\documentclass{iopjournal}
\usepackage{bm}
\usepackage{amsmath}
\usepackage{gensymb}
\usepackage{subcaption}
\usepackage{graphicx}
\usepackage{comment}

\begin{document}

\articletype{Paper} 

\title{A mobility based approach to transport in chiral fluids}

\author{Filippo Faedi $^1$, Erik Kalz $^2$,  Ralf Metzler $^2$, Abhinav Sharma$^{1,3,*}$}

\affil{$^1$University of Augsburg, Institute of Physics, D-86159 Augsburg, Germany}

\affil{$^2$University of Potsdam, Institute of Physics and Astronomy, D-14476 Potsdam, Germany}

\affil{$^3$Leibniz-Institute for Polymer Research, Institute Theory of Polymers, D-01069 Dresden, Germany}

\email{abhinav.sharma@uni-a.de}


\begin{abstract}
Chiral fluids, for which the mobility tensor has antisymmetric, off-diagonal components, exhibit transport phenomena absent in conventional systems, including interaction-enhanced diffusion and negative mobility. While these effects have been predicted theoretically and observed in simulations, their microscopic origin has remained unclear. Here, we address this question using a mobility-based nonequilibrium approach, analysing the steady-state drift of a tracer driven through an interacting chiral fluid. We show that, under strong chirality, the tracer generates a reversed density wake, in which regions of particle accumulation and depletion are inverted compared to the achiral case. This structural inversion of the wake provides a unified physical mechanism underlying both enhanced diffusion and negative mobility. Furthermore, we demonstrate that these phenomena are robust to changes in the interaction potential, highlighting their generality as a consequence of odd mobility.
\end{abstract}

\section{Introduction}
\label{sec_introduction}

In overdamped Brownian dynamics, particle motion is described at the microscopic level by a linear relation between forces and velocities. Specifically, the instantaneous particle velocity is proportional to the total force acting on it, 
\begin{equation}
\label{velocity_force_relation}
    \mathbf{v} = \boldsymbol{\mu} \ \mathbf{f}_{\mathrm{tot}},
\end{equation}
where $\mathbf{v}$ is the particle velocity, $\mathbf{f}_{\mathrm{tot}}$ is the total force acting on the particle and $\boldsymbol{\mu}$ is the mobility tensor. In ordinary Brownian systems, under conditions of isotropy and time-reversal symmetry, the mobility tensor is diagonal $\boldsymbol{\mu}= \mu_0 \mathbf{1}$, where $\mu_0$ denotes the bare mobility and $\mathbf{1}$ is the identity tensor. The total force $\mathbf{f}_{\mathrm{tot}}$ generally consists of three contributions: stochastic thermal forces arising from the surrounding heat bath, interaction forces due to other particles, and possible external driving forces.
This description neglects inertial effects, corresponding to the overdamped limit in which momentum relaxes on time scales much shorter than those of interest for the particle dynamics \cite{aakesson1985brownian}.

In contrast to ordinary Brownian systems, chiral fluids are characterised by a single-particle mobility tensor that contains an antisymmetric contribution \cite{reichhardt2019active, poggioli2023odd, kalz2025reversal}. This antisymmetric part breaks the mirror symmetry and leads to a coupling between forces and transverse motion. In two dimensions, the mobility tensor can be written as \cite{poggioli2023odd}
\begin{equation}
\label{odd_mobiity_tensor}
    \boldsymbol{\mu} = \mu_0 \left( \boldsymbol{1} + \kappa \boldsymbol{\epsilon} \right),
\end{equation}
where $\mu_0$ is the bare mobility as above, $\boldsymbol{\epsilon}$ denotes the antisymmetric Levi-Civita symbol, and $\kappa$ is a dimensionless parameter hereafter referred to as the oddness parameter. For $\kappa=0$, the antisymmetric contribution vanishes and one recovers the purely diagonal mobility of conventional Brownian particles. The antisymmetric term produces a velocity component perpendicular to the applied force, a hallmark signature of chiral transport. Such odd mobility is a generic feature of chiral systems and has been observed across a wide range of physical realisations, including magnetic skyrmions \cite{troncoso2014brownian, reichhardt2015collective, litzius2017skyrmion}, driven colloidal assemblies \cite{soni2019odd, massana2021arrested, cao2023memory}, and motile microorganisms \cite{diluzio2005escherichia, marcos2012bacterial, petroff2015fast}. Depending on the system, it may originate intrinsically from particle-level dynamics \cite{jiang2017direct, drescher2009dancing, jing2020chirality} or arise effectively through coupling to odd-viscous environments or external fields \cite{hargus2025odd, chun2018emergence, nelson2025topological, shinde2022strongly}.
Chiral fluids exhibit a number of striking and unconventional dynamical properties, reflecting a modification of how interactions affect 
transport. In conventional Brownian systems, interparticle interactions invariably hinder motion: collisions increase friction and reduce the long-time self-diffusion coefficient of the tracer, $D_\mathrm{s}$, below its short-time value $D_0$ \cite{royall2024colloidal}. 
This slowdown occurs independently of the nature of the interactions. For instance, in 
dilute hard-disk systems, one finds the classical result $D_\mathrm{s} = D_0 (1 - 2\phi)$, where 
$\phi$ is the area fraction, a result obtained using a variety of approaches, including direct calculations of transition probability densities \cite{hanna1981velocity} and mobility-based formalisms \cite{dhont1996introduction}. Remarkably, chiral systems invert this familiar paradigm. Previous studies showed that, in the presence of a mobility with the structure of Eq.~\eqref{odd_mobiity_tensor}, interactions can enhance rather than suppress self-diffusion, leading to $D_\mathrm{s} > D_0$ even for purely repulsive particles 
\cite{kalz2022collisions, kalz2024oscillatory, muzzeddu2025self}. 
While normal diffusive particles recoil from one another upon collision, chiral ones exhibit a characteristic mutual rolling motion: curved probability 
fluxes generated by the antisymmetric part of the mobility tensor allow particles to move around obstacles \cite{kalz2022diffusion, mecke2025obstacle}. 
Despite these quantitative predictions and their confirmation in simulations, the physical mechanism by which interactions are converted from a source of friction into a source of enhanced transport has remained largely unexplained.

In this work, we adopt a perspective based on mobility that is complementary to the earlier microscopic approach \cite{kalz2022collisions, kalz2024oscillatory}. We consider a tracer particle driven through a chiral fluid by a constant external force, and we compute its steady-state drift velocity in the presence of interactions with the surrounding particles. 
The relation between the steady state velocity and the applied force defines an effective mobility tensor which includes the emerging effects of the interactions with the host particles and which turns out to have the same antisymmetric structure. Via the fluctuation-dissipation relation, this effective mobility also determines the long-time diffusion tensor of the tracer, thereby establishing a direct connection between nonequilibrium response and spontaneous fluctuations.

This mobility-based approach not only confirms previous findings on the dependence of self-diffusion on the oddness parameter $\kappa$ 
\cite{kalz2022collisions, kalz2024oscillatory}, but also provides a transparent physical and unified interpretation of their origin. Solving the explicitly out-of-equilibrium driven problem allows us to characterise the density disturbance induced by the tracer in the surrounding fluid. For repulsive interactions and small $\kappa$, particles accumulate in front of the tracer and are depleted behind it, increasing the effective drag in an intuitive way. As $\kappa$ increases, however, this pattern undergoes a qualitative inversion: depletion shifts to the front while accumulation forms behind the tracer. As a result, the friction experienced by the tracer is reduced.

This inversion of the interaction-induced density field provides a simple and intuitive 
mechanism for interaction-enhanced diffusion in chiral fluids. The same physical picture 
also naturally explains the emergence of negative mobility in driven chiral fluids \cite{kalz2025reversal}: once the force-induced 
distortion of the surrounding medium becomes sufficiently inverted, the net interaction 
force exerted by the host particles can overcome the direct response to the applied force, 
causing the tracer to drift opposite to the applied drive. We further extend our analytical framework to attractive interactions between particles. 
In this case, we find that the effect of interactions on self-diffusion is likewise reversed 
in the limit of large oddness $\kappa$, highlighting once again how odd mobility fundamentally 
reshapes the interplay between forces, correlations, and transport in chiral fluids, independent of the specific interaction details.

In Sec. \ref{sec_dynamics}, we formulate the dynamics of interacting chiral Brownian particles at the level of the many-body probability density. This leads to a configurational evolution equation that generalises the Smoluchowski equation by incorporating an antisymmetric (odd) mobility.
In Sec. \ref{subsec_steady_state}, we derive a relation between the effective tracer mobility and the stationary density wake generated in the surrounding medium. We then obtain an analytical solution for this density wake for hard-disk interactions in Sec. \ref{subsec_hard_interactions}, and use it to analyse the resulting phenomena of enhanced self-diffusion and negative mobility.
Finally, in Sec. \ref{subsec_attractive_interactions}, we extend the theory of effective mobility to systems where short-range attractive interactions are superimposed on hard-disk repulsion. In Sec. \ref{sec_conclusions} we give an outlook and conclude.

\section{Dynamics of interacting chiral Brownian particles}
\label{sec_dynamics}

We consider a two-dimensional fluid composed of $N$ interacting chiral Brownian particles that are in thermal equilibrium with a solvent at temperature $T$. The single-particle mobility tensor is assumed to have the form given in Eq.~\eqref{odd_mobiity_tensor}. Each particle experiences an interaction force $\mathbf{f}_{i, \mathrm{int}}=-\nabla_i U_N$ arising from the $N$-body interaction potential $U_N$, where $\nabla_i=\partial/\partial \mathbf{r}_i$ is the partial derivative with respect to particle $i$'s position $\mathbf{r}_i$, and a thermal force $\mathbf{f}_{i, \mathrm{th}} $ due to the surrounding solvent. According to the force-velocity relation in Eq.~\eqref{velocity_force_relation} the velocity of particle $i$ can therefore be written as

\begin{equation}
\label{moiblity}
     \mathbf{v}_i = \boldsymbol{\mu} \left[ \mathbf{f}_{i, \mathrm{int}}   + \mathbf{f}_{i, \mathrm{th}} \right].
\end{equation}

Given the single-particle velocities, the time evolution of the many-body probability density can be obtained from a continuity equation in configurational space. Following the standard derivation of Ref.~\cite{dhont1996introduction}, this yields 

\begin{equation}
\label{continuity}
    \frac{\partial}{\partial t} P_N(\{ \mathbf{r}_j\}, t) = - \sum_{i=1}^N \nabla_i \cdot   P_N(\{ \mathbf{r}_j\}, t) \boldsymbol{\mu} \left[ \mathbf{f}_{i, \mathrm{t}} - \nabla_i U_N \right],
\end{equation}
where $P_N(\{ \mathbf{r}_j\}, t)$ is the joint probability density function (PDF) to find the $N$ particles in the spatial configuration $\{ \mathbf{r}_j\} = \{ \mathbf{r}_1,\ldots,  \mathbf{r}_N$ \} at time $t$. Imposing that the steady state is an equilibrium Boltzmann distribution with zero probability current determines the form of the thermal force as

\begin{equation}
    \label{thermal_force}
     \mathbf{f}_{i, \mathrm{th}}=-k_B T \nabla_i \ln P_N(\{ \mathbf{r}_j\}, t).
\end{equation}

This corresponds to the Einstein relation, expressing the fluctuation–dissipation balance between thermal noise and dissipative mobility. Substituting this result back into the continuity equation yields the Smoluchowski equation governing the dynamics of the chiral particle system,

\begin{equation}
\label{N_body_Smoluchowski_eq}
    \frac{\partial}{\partial t} P_N(\{\mathbf{r}_j\}, t) = \sum_{i=1}^N \nabla_i \cdot  \boldsymbol{D}[\nabla_i + \beta \nabla_i U_N(\{\mathbf{r}_j\}) ] P_N(\{\mathbf{r}_j\}, t),
\end{equation}
where $\beta=1/k_\mathrm{B} T$ and $\boldsymbol{D}$ is the single particle diffusion tensor, related to the mobility through the Einstein relation $\boldsymbol{D} = k_\mathrm{B}T\, \boldsymbol{\mu}$. As a consequence, the diffusion tensor inherits both symmetric and antisymmetric components from the mobility.
The Smoluchowski equation \textcolor{red}{\eqref{N_body_Smoluchowski_eq}} can also be derived formally from the full phase-space dynamics of Brownian particles subject to a Lorentz force, as shown in Ref.~\cite{chun2018emergence}.
Notably, the antisymmetric (chiral) part of the diffusive flux is divergence-free. As a result, it does not affect the diffusion of a free particle, which remains identical to that of ordinary Brownian motion. Its effects become manifest only in the presence of constraints such as reflecting boundary conditions, for instance those arising from hard-disk interactions.

\subsection{Steady-state pair distribution around a driven tracer}
\label{subsec_steady_state}

We now consider the system introduced above and focus on the steady-state velocity of a tracer particle driven by a constant external force $\mathbf{f}_\mathrm{ext}$. The tracer, labelled ``1'', interacts with $N$ identical particles referred to as host particles in a confined volume $V$. Determining the tracer velocity requires computing the nonequilibrium pair (radial) distribution function, which encodes the stationary density distortion generated around the driven particle. This quantity is crucial because it directly determines the average interaction force exerted by the host particles on the tracer. In the following, we adopt the classical treatment developed by Dhont for ordinary Brownian suspensions \cite{dhont1996introduction}. We further assume pairwise additive interactions of the form $U_N(\{\mathbf{r}_j\}) = \sum_{i=1}^N \sum_{ j=i+1}^{N} U(r_{ij})$, where $r_{ij} = |\mathbf{r}_i - \mathbf{r}_j|$, and focus on the dilute regime. In this limit, many-body collisions are negligible, and the problem reduces to solving the two-body Smoluchowski equation
\begin{equation}
\label{2_body_Smoluchowski_eq}
    \frac{\partial}{\partial t} P_2(\mathbf{r}_1, \mathbf{r}_2, t) = \nabla_1 \cdot \boldsymbol{D}[ \nabla_1 +  \beta\nabla_1 U(r) - \beta\mathbf{f}_\mathrm{ext}]P_2(\mathbf{r}_1, \mathbf{r}_2, t)  + \nabla_2 \cdot \boldsymbol{D}[  \nabla_2 +  \beta
    \nabla_2 U(r)]P_2(\mathbf{r}_1, \mathbf{r}_2, t).
\end{equation}

The steady-state solution of this equation differs qualitatively from Eq.~\eqref{N_body_Smoluchowski_eq} due to the presence of the external driving force, which generates nonvanishing probability currents. As a result, the stationary distribution is not of Boltzmann form. The solution can be factorised into a uniform centre-of-mass contribution and an internal part depending only on the relative coordinate, $P_2^\mathrm{ss}(\mathbf{r}_1, \mathbf{r}_2) = P_\mathrm{com}^\mathrm{ss}(\mathbf{R})\, P_\mathrm{rel}^\mathrm{ss}(\mathbf{r})$ where $\mathbf{R} = (\mathbf{r}_1 + \mathbf{r}_2)/2$ is the centre of mass and $\mathbf{r} = \mathbf{r}_2 - \mathbf{r}_1$ is the relative particle coordinate. The centre of mass distribution is uniform, $P_\mathrm{com}^\mathrm{ss}(\mathbf{R})=1/V$, while the internal part satisfies the following equation 
\begin{equation}
\label{inner_equation}
\nabla_\mathbf{r} \cdot \boldsymbol{D} \left[ \nabla_\mathbf{r} + \beta ( \nabla_\mathbf{r} U(r) + \mathbf{f}_\mathrm{ext} /2)  \right]  P_\mathrm{rel}(\mathbf{r}) =0,
\end{equation}
which can be solved independently.

The relative steady-state distribution can be expressed as a perturbation expansion in the strength of the external force:
\begin{align}
\label{perturbative_g_ansatz}
P_\mathrm{rel}^\mathrm{ss}(\mathbf{r})=\frac{g_\mathrm{eq}(\mathbf{x})}{V}\left(1+ \mathrm{Pe}\, g(\mathbf{x}) + \mathcal{O}(\mathrm{Pe}^2)\right),
\end{align}
where $\mathbf{x} = \mathbf{r} / \sigma$ is the dimensionless relative coordinate, $\sigma$ denotes the particle diametre and  $\mathrm{Pe} = \sigma \beta|\mathbf{f}_\mathrm{ext}|/2$ is the P{\'e}clet number. The equilibrium radial distribution function $g_\mathrm{eq}(\mathbf{x})$ in the dilute limit is given by the Boltzmann function 
\begin{align}
\label{equilibrium_radial_function}
g_\mathrm{eq}(\mathbf{x}) = \mathrm{exp}(-\beta U(x)),.
\end{align}
and thus only is a function of the radial distance $x=|\mathbf{x}|$ for an isotropic interaction potential $U(x)$. 
The function $g(\mathbf{x})$ in Eq.~\eqref{perturbative_g_ansatz} represents the nonequilibrium correction to the radial distribution induced by the external drift. Unlike the equilibrium distribution, $g(\mathbf{x})$ can be anisotropic. 
Moreover, the spatial integral of the second term should vanish to ensure correct normalisation of $P_\mathrm{rel}^\mathrm{ss} $. For chiral fluids, the solution deviates from the classical result obtained by Dhont for ordinary Brownian particles \cite{dhont1996introduction}: the assumption of symmetry around the driving direction is relaxed due to the antisymmetric component of the mobility tensor.
Inserting the perturbative ansatz into Eq.~\eqref{inner_equation} yields an equation for $g(\mathbf{x})$. While the zeroth-order solution in $\mathrm{Pe}$ is already known, the first-order equation reads
\begin{align}
\label{equation_for_distorted_g}
    \nabla_\mathbf{x} \cdot \boldsymbol{D}\left[g_\mathrm{eq}(x)\left(\hat{\mathbf{e}}_{f} + \nabla_\mathbf{x} g(\mathbf{x})\right)\right] &= 0,
\end{align}
where $\hat{\mathbf{e}}_{f} = \mathbf{f}_\mathrm{ext} / |\mathbf{f}_\mathrm{ext}|$ is the unit vector along the external force.
The solution of Eq.~\eqref{equation_for_distorted_g} depends on the specific form of the interaction potential through $g_\mathrm{eq}$,
and will be analyzed in detail in the following sections for two representative cases.

\subsection{Hard-disk interactions: Enhanced self-diffusion and negative mobility}
\label{subsec_hard_interactions}

We now consider a dilute system of chiral Brownian particles interacting via hard-disk potentials, for which $\sigma$ is the actual particle diameter. In this limit, the equilibrium radial distribution function simplifies to a Heaviside function at the contact distance $x=1$, i.e. $r=\sigma$,
\begin{equation}\label{eq:_geq_HS}
\begin{split}
g_\mathrm{eq}(x)=\Theta(x-1). 
\end{split}
\end{equation}
Since $g_\mathrm{eq}$ is a constant in the domain $x \geq 1$, the first term in Eq.~\eqref{equation_for_distorted_g} vanishes and the nonequilibrium correction $g(\mathbf{x})$ satisfies the Laplace equation

\begin{equation}\label{laplace_HS}
\Delta_x g(\mathbf{x})=0 \, \,\text{for} \,  \,  x \geq 1.
\end{equation}

\begin{figure}
\centering
    \includegraphics[width=\textwidth]{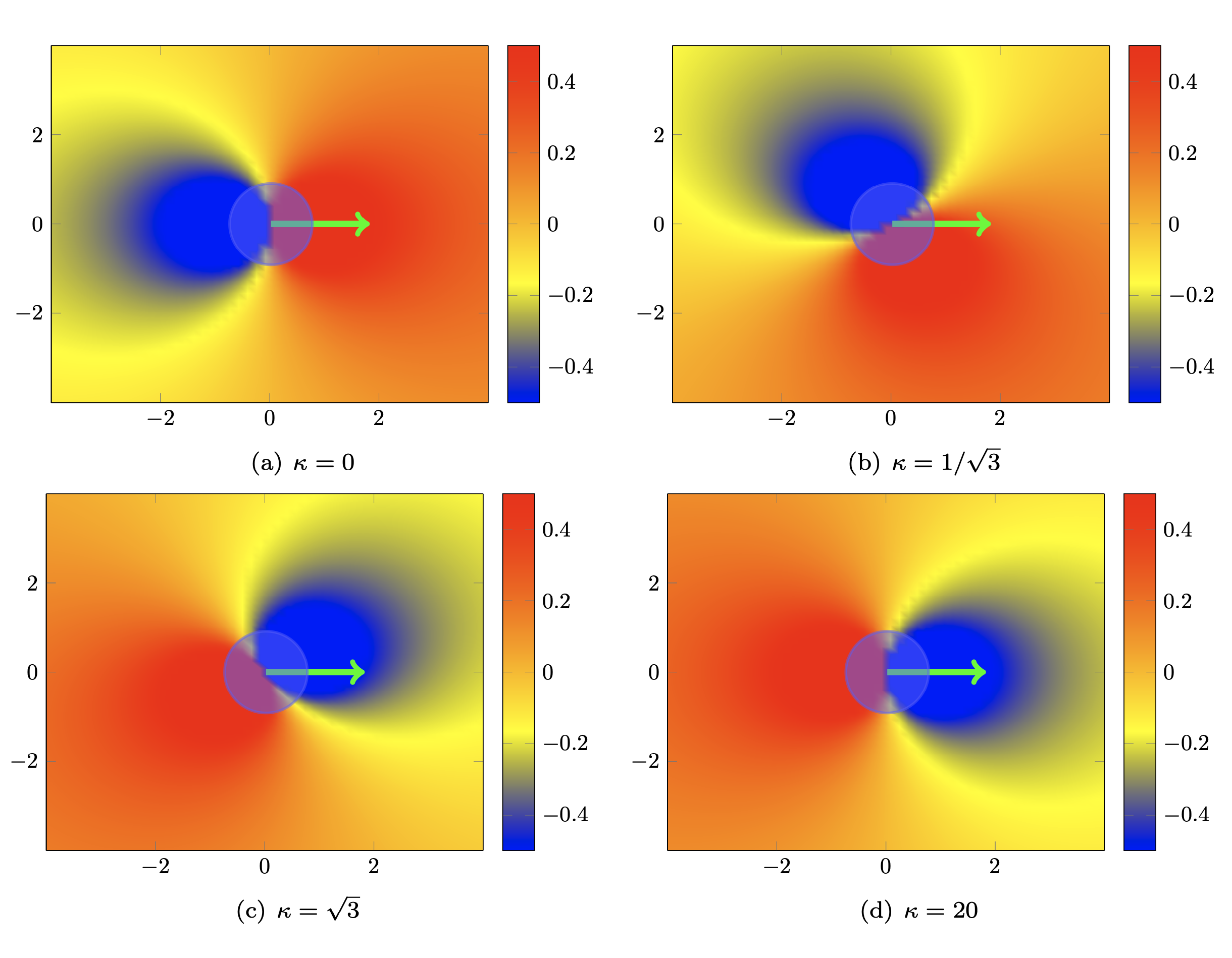}
    \caption{Colour map of the particle wake generated by a tracer particle in the $(x,y)$ plane for different values of the oddness parameter $\kappa$ at area fraction $\phi = 0.1$. Panels correspond to (a) $\kappa = 0$, (b) $\kappa = 1/\sqrt{3}$, (c) $\kappa = \sqrt{3}$, and (d) $\kappa = 20$. Axes are scaled in units of the particle diametre. In (a), particles accumulate in front of the tracer and are depleted behind it relative to the driving force (green arrow). As $\kappa$ increases, this anisotropy gradually rotates, reaching $\pi$ rotation in the high chirality limit $\kappa\to \infty$. In this regime, collisions occur preferentially from behind, effectively propelling the tracer forward.}
    \label{fig:ne_rdf}
\end{figure}

To determine the solution, we impose reflecting boundary conditions at contact, generated by the hard-disk potential. Specifically, the relative particle flux
\begin{equation}
    \label{relative_particle_flux}
    \mathbf{J}_{r}= \ g_\mathrm{eq}(x)\boldsymbol{D}\left(\mathbf{e}_{f} - \nabla_\mathbf{x} g(\mathbf{x})\right).
\end{equation}
must satisfy 
\begin{equation}
    \label{reflecting_BC}
    \hat{\mathbf{x}} \cdot \mathbf{J}_{r} \vert_{x=1} = 0,
\end{equation}
where $\hat{\mathbf{x}}=\mathbf{x}/x$ is the unit vector along the radial distance. This condition leads to the coupled equations
\begin{subequations}\label{eq:_g_x_HS}
\begin{align}
 &\Delta_x g(\mathbf{x})=0, \, \,\text{for} \,  \,  x\geq 1,\label{eq:_g_x_HSa}\\ 
 &\hat{\mathbf{x}} \cdot \boldsymbol{D}\hat{\mathbf{e}}_f= - \hat{\mathbf{x}} \cdot  \boldsymbol{D} \nabla_x g(\mathbf{x}), \, \,\text{at} \,  \,  x=1, \label{eq:_g_x_HSb}
\end{align}
\end{subequations}
which can be solved simultaneously in $g(\mathbf{x}) $ imposing the additional condition that $g(\mathbf{x})$ vanishes as $x \rightarrow \infty$. 
The unique solution is 

\begin{equation}
\label{solution}
g(\mathbf{x})=\hat{\mathbf{a}}\cdot\hat{\mathbf{x}}/x,
\end{equation}
where the unit vector $\hat{\mathbf{a}}$ encodes the anisotropy induced by the antisymmetric mobility
\begin{subequations}
\begin{align}
\label{unit_vector_HS}
\hat{\mathbf{a}}&=\frac{\boldsymbol{D}^2}{1+\kappa^2} \hat{\mathbf{e}}_{f}=\left[ a_{\parallel} \mathbf{1} + a_{\perp} \boldsymbol{\epsilon} \right] \cdot \hat{\mathbf{e}}_{f}\\
 a_{\parallel} &= \frac{1-\kappa^2}{1+\kappa^2},\  a_{\perp} = \frac{2\kappa}{1+\kappa^2}
\end{align}
\end{subequations}
The solution $g(\mathbf{x})$ is plotted in Fig.~\ref{fig:ne_rdf}, where we observe that for $\kappa=0$ (no chirality), the nonequilibrium distribution, which has a structure of a density wake from the perspective of the driven tracer, is aligned with the external driving direction, while increasing $\kappa$
rotates the wake continuously, completing a $\pi$ rotation in the strong-chirality limit $\kappa>>1$.

The effective mobility of the tracer is determined by the average interaction force it experiences due to the host particles:
\begin{equation}
\label{ave_inter_force}
\begin{split}
   \mathbf{f}_\mathrm{int}=-\left<\nabla_1 U_N\right>,
\end{split}
\end{equation}
where the average is taken over the steady-state distribution. In the dilute limit, each host particle contributes independently, giving
 \begin{equation}
\label{ave_inter_force_1}
\mathbf{f}_\mathrm{int}=N \int d\mathbf{r} P_\mathrm{rel}^{\mathrm{ss}}(\mathbf{r}) \nabla_r U(r) 
    =     - 2\phi \frac{\boldsymbol{D}^2}{1+\kappa}\ \mathbf{f}_\mathrm{ext}     
\end{equation}
where $\phi=N \pi \sigma^2/4V$ is the area fraction of the particles. The second equality follows from the derivative of the Heaviside function, as detailed in Appendix \ref{app_A}. Since the average thermal force on the tracer vanishes, the total average force acting on the tracer is
\begin{equation}
\label{ave_tot_force}
\begin{split}
   \mathbf{f}_\mathrm{tot}  = \left[\mathbf{1}  - 2\phi \frac{\boldsymbol{D}^2}{1+\kappa}\right] \mathbf{f}_\mathrm{ext} 
\end{split}
\end{equation}

The tracer velocity follows from the single-particle mobility, leading to an expression for the effective mobility,
\begin{subequations}
\begin{equation}\label{mu_eff_HS}
\mathbf{v}= \boldsymbol{\mu} \left[\mathbf{1}  - 2\phi \frac{\boldsymbol{D}^2}{1+\kappa} \right]\mathbf{f}_\mathrm{ext} =\beta\left[ D_{\parallel}\mathbf{1}+\kappa D_{\perp}\boldsymbol{\epsilon} \right]\mathbf{f}_\mathrm{ext},
\end{equation}
with 
\begin{equation}
\label{diagonal_D}
D_{\parallel}=D_0\left( 1-2\phi \frac{1-3\kappa^2}{1+\kappa^2}\right)
\end{equation}
and
\begin{equation}
\label{offdiagonal_D}
D_{\perp}=D_0\left( 1-2\phi \frac{3-\kappa^2}{1+\kappa^2}\right).
\end{equation}
\end{subequations}
The effective mobility preserves the symmetry of the single-particle tensor: it has an antisymmetric component $D_{\perp}$
and a diagonal part $D_{\parallel}$. Applying the Einstein relation then gives the long-time self-diffusion tensor as

\begin{equation}\label{D_self}
\begin{split}
\boldsymbol{D}_s=\left[ D_{\parallel}\mathbf{1}+\kappa D_{\perp}\boldsymbol{\epsilon} \right].
\end{split}
\end{equation}
Here, the diagonal elements $D_{\parallel}$ corresponds to the previously obtained result for the self-diffusion coefficient in Refs.~\cite{kalz2022collisions} and \cite{kalz2024oscillatory}.

\begin{figure}
\centering
\begin{subfigure}{0.45\textwidth}
    \includegraphics[width=\textwidth]{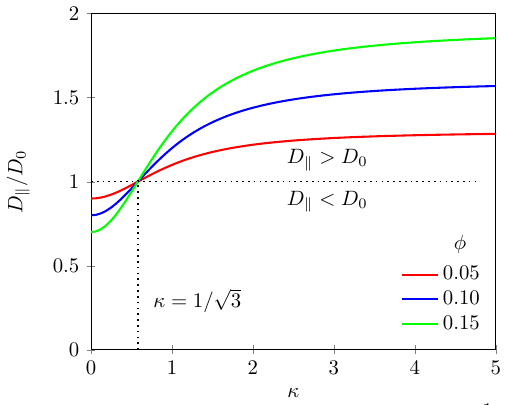}
    \caption{}
    \label{fig:first}
\end{subfigure}
\hfill
\begin{subfigure}{0.45\textwidth}
    \includegraphics[width=\textwidth]{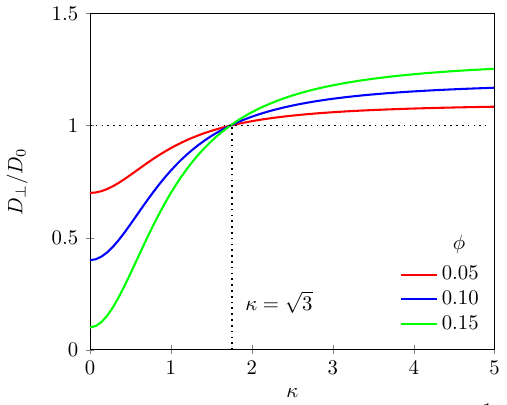}
    \caption{}
    \label{fig:second}
\end{subfigure}    
\caption{(a) Reduced self-diffusion coefficient $D_{\parallel}/D_0$ as a function of the odd-parameter $\kappa$ and the host area fraction $\phi$. For $\kappa \to \infty $, the diffusion coefficient approaches $D_\parallel=D_0(1+6\phi)$ whereas for $\kappa=0$, it recovers the known hard-disk result $D_\parallel=D_0(1-2\phi)$.(b) $D_{\perp}/D_0$ as a function of $\kappa$ and $\phi$. }
\label{fig:foobar}
\end{figure}

The diagonal and off-diagonal components of the mobility tensor, $D_{\parallel}$ and $D_{\perp}$ of Eqs.~\eqref{diagonal_D} and \eqref{offdiagonal_D} are shown in Fig.~\ref{fig:foobar} as a function of oddness parameter $\kappa$ for different densities $\phi$ in the dilute regime. Notably, the special values 
$\kappa=1/\sqrt3$ and $\kappa=\sqrt{3}$ render the diagonal and off-diagonal components of the self-diffusion tensor, respectively,  independent of concentration. In particular, for $\kappa=1/\sqrt3$ the tracer’s diffusion equals that of a free particle, while for $k>1/\sqrt3$ it becomes enhanced. In the extreme limit 
$\kappa \to \infty$, the self-diffusion reduces to 

\begin{equation}\label{eq:_D_enhanced}
    \lim_{\kappa \to \infty} D_{\parallel} =D_0(1+6\phi),
\end{equation}
and thus shows a transition from interaction reduction to interaction enhancement, as can be seen in Fig.~\ref{fig:foobar}

The enhancement of tracer dynamics can be understood by examining how the nonequilibrium radial distribution function depends on the oddness parameter $\kappa$, as shown in Fig.~\ref{fig:ne_rdf}. For $\kappa=0$, the external force produces an accumulation of host particles in front of the tracer and a depletion behind it, relative to the direction of motion. This front-back asymmetry increases the frequency of collisions from the front, thereby reducing the tracer mobility. As $\kappa$ increases, the anisotropy of the distribution gradually rotates, reaching a full $\pi$-rotation in the limit $\kappa \to \infty$. In this regime, the tracer wake shifts predominantly behind the particle, creating an effective propulsion: the average interaction force now points opposite to the external drive, enhancing forward motion.

This mechanism of inversion of the average effect of the interaction force not only affects the tracer particle's diffusion. It also affects its drift-response on an external force, to such an extent that the tracer particle in a chiral fluid can exhibit absolute negative mobility \cite{kalz2025reversal}. Absolute negative mobility thereby is the effect of an average velocity being opposite to the direction of an externally applied force \cite{eichhorn2002brownian, eichhorn2002paradoxical}. When both the tracer and the host particles are driven in the same direction, this effect can even arise together with a reversal of the intrinsic Hall-drift of a chiral tracer  \cite{kalz2025reversal}. This effect can be understood phenomenologically by considering the behaviour of the nonequilibrium radial distribution function, which depends only on the relative forces acting on the particles. Following the derivation of Sec. \ref{subsec_hard_interactions}, we replace the external force in Eq.~\eqref{inner_equation} with an effective force
\begin{equation}
    \label{effective_force}
    \mathbf{f}_\mathrm{eff}=\mathbf{f}_1-\mathbf{f}_2=(1-\omega)\mathbf{f}_1,
\end{equation}
where $\mathbf{f}_1$ and $\mathbf{f}_2$ are the forces acting on the tracer and the host particle, respectively, and $\omega = |\mathbf{f}_2|/|\mathbf{f}_1|$. 
When $|\mathbf{f}_2|>|\mathbf{f}_1|$, i.e. $\omega >1$, the tracer wake reverses with $\kappa$: for 
$\kappa=0$, the distribution resembles that shown in Fig.~\ref{fig:ne_rdf}(d), aligned with $\mathbf{f}_1$, whereas in the limit $\kappa \to \infty$, the wake rotates to the configuration of Fig.~\ref{fig:ne_rdf}(a). Consequently, without chirality, the average interaction force enhances the motion along the driving direction, as the tracer particle is driven with a stronger force than the host particles. In contrast, in the strong-chirality limit, $\kappa \to \infty$, interactions with host particles oppose the applied force, potentially reversing the tracer's velocity. In the opposite case, $|\mathbf{f}_2|<|\mathbf{f}_1|$, i.e. $\omega <1$, the tracer particle is driven with a weaker force than the host particles and the $\kappa=0$ distribution resembles that of Fig.~\ref{fig:ne_rdf}(a), aligned with $\mathbf{f}_1$. For the strong chirality limit, the wake rotates to the configuration of Fig.~\ref{fig:ne_rdf}(d) and the tracer particles' motion is enhanced by collisions with the host particles. Following the perturbative derivation described earlier, the steady-state tracer velocity can be expressed as
\begin{equation}
    \mathbf{v}_1 =\beta \boldsymbol{D} \left(\mathbf{1} - 2\phi (a_{\parallel} \mathbf{1} + a_{\perp} \boldsymbol{\epsilon})(1-\omega)\right)\mathbf{f}_1,
\end{equation}
where $a_{\parallel}$ and $a_{\perp}$ are given by Eq.~\eqref{unit_vector_HS} for hard-disk interactions. Decomposing the velocity into components parallel and perpendicular to $\mathbf{f}_1$, $\mathbf{v}_1=(v_{\parallel}, v_{\perp})$, gives

\begin{equation}
\label{v_parr}
    v_{\parallel}/(\beta D_0 |\mathbf{f}_1|)= 1 -2\phi \frac{1-3\kappa^2}{1+\kappa^2}
    (1 -\omega)
\end{equation}
and
\begin{equation}
    v_{\perp}/(\beta D_0 |\mathbf{f}_1|)= 1 - 2\phi \frac{3-\kappa^2}{1+\kappa^2}
   (1 -\omega).
\end{equation}

\begin{figure}[h]
\centering
\begin{subfigure}{0.45\textwidth}
    \includegraphics[width=\textwidth]{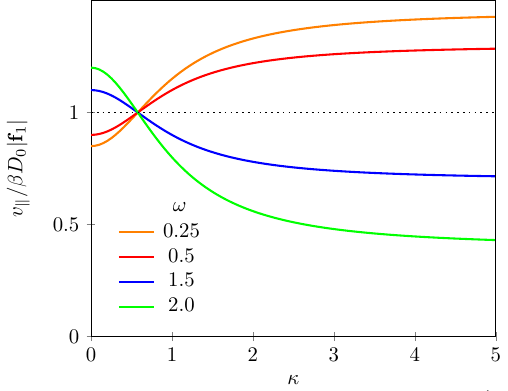}
    \caption{}
    \label{fig:first}
\end{subfigure}
\hfill
\begin{subfigure}{0.45\textwidth}
    \includegraphics[width=\textwidth]{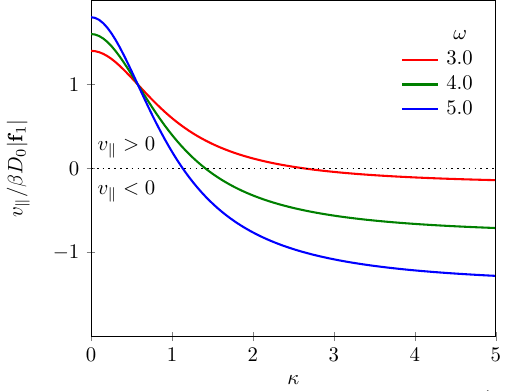}
    \caption{}
    \label{fig:second}
\end{subfigure}    
\caption{Reduced parallel tracer velocity $v_{\parallel} /(\beta D_0 |\mathbf{f}_1|)$   as a function of the oddness parameter $\kappa$ for a a volume fraction $\phi=0.1$. (a) Values of the drift ratio  $\omega$ such that the motion of the tracer particle is either enhanced or reduced by interactions with the host particles. (b) Values of $\omega$ such that the negative mobility condition $6\phi (\omega - 1) > 1$ is satisfied, and the parallel tracer velocity becomes negative as a function of oddness $\kappa$.}
\label{fig:omega}
\end{figure}

From Eq.~\eqref{v_parr}, the condition for absolute negative mobility along the driving direction ($\left<v_{\parallel}\right><0$) is $6\phi (\omega - 1) > 1$.
For given values of $\omega$ and $\phi$ satisfying this inequality, the tracer velocity reverses for odd parameters $\kappa$larger than
\begin{equation}
    \kappa \ge \frac{1+2\phi(\omega-1)}{6 \phi (\omega-1) - 1}.
\end{equation}
The value of the parallel component of the velocity is plotted in Fig.~\ref{fig:omega} as a function of $\omega$ and $\kappa$.
This analytical result reproduces the findings of Ref.~\cite{kalz2025reversal}, which were obtained through an independent derivation. Numerical simulations in that work confirmed the validity of this prediction under dilute conditions.

\subsection{Effect of short-range attractions on tracer dynamics}
\label{subsec_attractive_interactions}

\begin{figure}[h]
    \centering
    \includegraphics[width=0.5\linewidth]{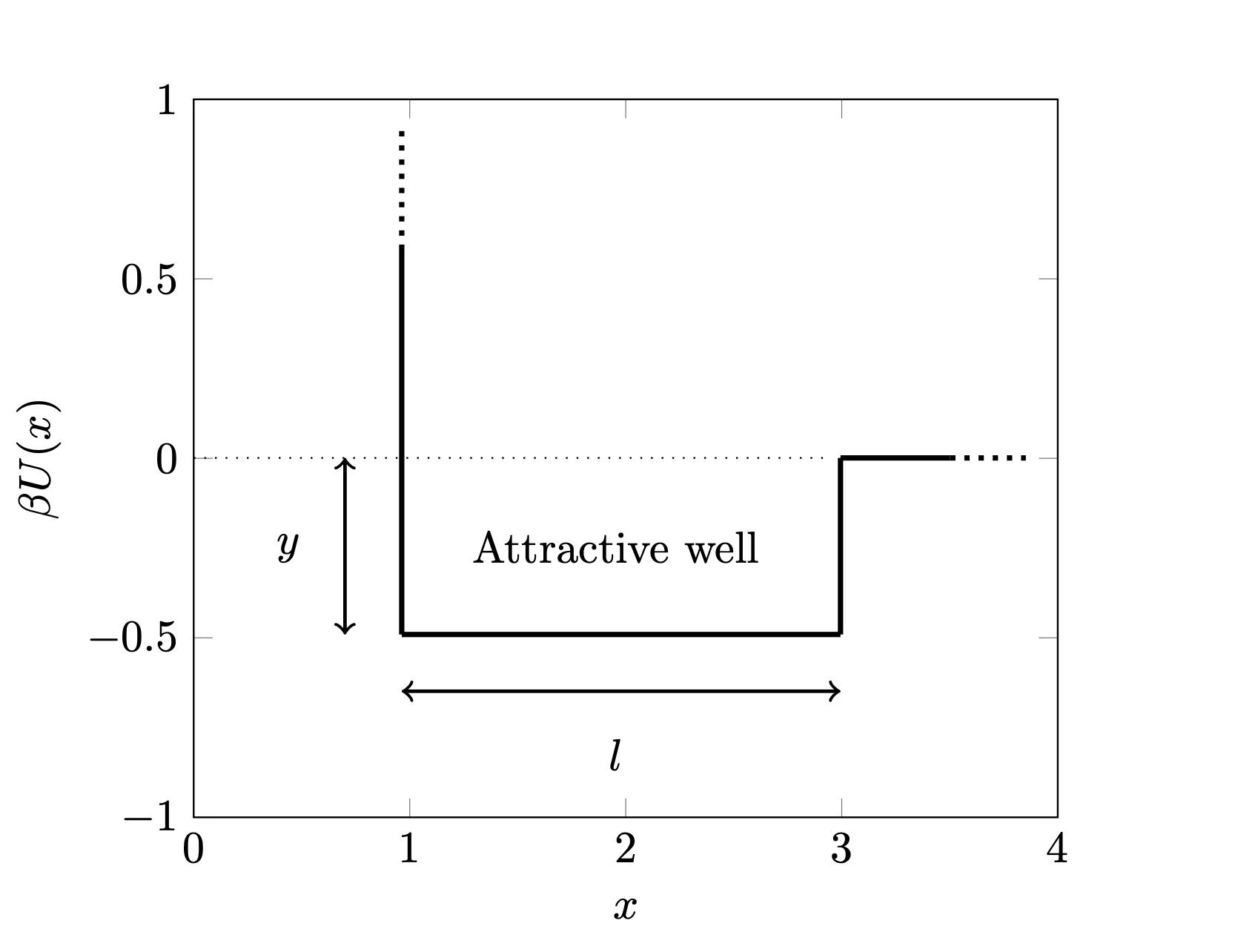}
\caption{Schematic of the weak attractive potential around a tracer particle. The potential is defined by its range $l$ and depth $y$. Particles are excluded from the hard-core region ($x <1$), experience a constant attraction in the intermediate region ($ 1 \le x < l+1 $), and diffuse freely for $x > l+1$.}
\label{fig:_attractive_potential}
\end{figure}

Analytical results for the effects of attractive interactions in chiral fluids have not been reported so far. In this section, we present a derivation of the nonequilibrium dynamics for particles interacting via a short-range attractive potential superimposed on a hard-disk core that is sketched in Fig.~\ref{fig:_attractive_potential}. We apply the method of effective mobility to study the self-diffusion of a tracer particle in this setting, following the approach discussed in Ref.~\cite{van1985exact}. The interaction potential around the tracer is divided into three distinct regions, as illustrated in Fig.~\ref{fig:_attractive_potential}: (1) an inner region ($x<1$), where hard-core exclusion prevents particle overlap, (2) an attractive well ($1 \leq x < l+1$), where the potential takes the constant value $U/(k_BT)=-y$ that represents the short-range attraction, and (3) an outer region ($x > l+1$), where the potential vanishes.

Due to the piecewise structure of the interaction potential, the equilibrium radial distribution function in the dilute limit can be written as a sum of two Heaviside functions, located at $x=1$ and $x=1+l$ as
\begin{equation}\label{eq:_g_eq_att}
\begin{split}
g_\mathrm{eq}(x)=\mathrm{e}^y\Theta(x-1)+(1-\mathrm{e}^y)\Theta(x-(l+1)).
\end{split}
\end{equation}

In both the attractive well and the outer region, the defining equation for the first-order correction to 
$g_\mathrm{eq}$, Eq.~\eqref{equation_for_distorted_g}, reduces to a Laplace equation for 
$g(\mathbf{x})$, since, similar to the hard-disk case, the divergence of the first term vanishes. The discontinuity of the potential at the boundaries naturally leads to a piecewise solution, defined by two distinct functions. We obtain

\begin{equation}
\label{sol_attr}
    g(\mathbf{x}) = \begin{cases}
       \mathbf{b} \cdot \mathbf{x} + (\mathbf{c} \cdot \hat{\mathbf{x}})/x, & \text{for} \, 1 \leq x < l+1 \\
       (\mathbf{d}\cdot \hat{\mathbf{x}})/x, & \text{for} \ x>l+1
     \end{cases}
\end{equation}
where $\mathbf{b}$, $\mathbf{c}$ and $\mathbf{d}$ are vectors determined via boundary conditions. The detailed calculations are presented in Appendix \ref{app_B}. Assuming a shallow attractive well, $y \ll \beta U$ , the average interaction force acquires a small correction to the result obtained for the pure hard-disk case and can be written as
\begin{equation}
\label{effective_mobility_micro_att}
   \mathbf{f}_\mathrm{int}= - 2\phi \frac{\boldsymbol{D}^2}{1+\kappa^2}\left[1-\gamma\right] \mathbf{f}_\mathrm{ext},
\end{equation}
where
\[\gamma=y \frac{(l+1)^2-1}{(l+1)^2}\]
is always positive. Using the same procedure as in the hard-disk case, the total average interaction force can be evaluated, and we find that, in the presence of the attractive well, the effective mobility defined in Eq.~\eqref{mu_eff_HS} takes the form

\begin{subequations}
\begin{align}
D_{\parallel}=D_0\left( 1-\phi \alpha  \frac{1-3\kappa^2}{1+\kappa^2}\right) \label{eq:_mu_eff_att_a},\\ 
D_{\perp}=D_0\left( 1-\phi \alpha  \frac{3-\kappa^2}{1+\kappa^2}\right), \label{eq:_mu_eff_att_b}
\end{align}
with
\begin{equation}
\alpha=2\left(1-y\frac{(l+1)^2-1}{(l+1)^2}\right).
\end{equation}
\end{subequations}

We find that the concentration dependence of the self-diffusion coefficient can be factorised into two contributions: a parameter 
$\alpha$, which encodes the effect of the interaction potential, and a chirality-dependent term that drives a sign-change of the interaction-correction. The latter is identical to the hard-disk case, while $\alpha$ depends explicitly on both the depth $y$ and the range $l$ of the attractive well. Consistency with the hard-disk result is recovered by setting either $y=0$ or $l=0$.
The validity of this expression is limited to $0<\gamma < 1$, as it is derived from a perturbative expansion around a vanishing attractive depth. Importantly, although the calculations presented here focus on self-diffusion in the presence of an attractive well, the qualitative behaviour is the same as in hard-disk systems. The effects arising from the antisymmetric (odd) component of the mobility—such as enhanced diffusion and negative mobility—are inherent to odd mobility. The results can be easily generalised to predict that negative mobility will also occur for attractive potentials, demonstrating the universality of these phenomena across different interaction types.

\section{Discussion and Conclusions}
\label{sec_conclusions}
In this work, we have developed a mobility-based framework to analyse self-diffusion in chiral systems, providing a systematic route to connect microscopic odd mobility to macroscopic transport properties. By considering an explicitly nonequilibrium setting with a driven tracer, we uncover transport phenomena that have no counterpart in achiral media. We showed that strong chirality qualitatively restructures the surrounding colloidal density field generating a reversed density wake of host particles in the high-chirality regime. This inversion underlies both the enhancement of tracer self-diffusion and the emergence of negative mobility. Importantly, these effects are not limited to hard-disk interactions: we demonstrate that they persist in the presence of short-range attractive potentials, indicating that they are generic features of odd mobility, independent of the specific interaction details.
Our results clarify the physical mechanism behind these phenomena and establish density-wake formation as a unifying principle governing transport in chiral active and Brownian matter.

%Extending this framework to dense systems, where many-body correlations become essential, and to more complex interaction potentials represents a natural and promising direction for future theoretical and computational investigations.
Our results suggest several implications for dense interacting systems, where many-body correlations become essential. Since diffusion enhancement arises from antisymmetric interaction-induced couplings without increasing thermal noise, odd mobility provides a mechanism to modify relaxation timescales without changing equilibrium fluctuations. In dense suspensions and crystalline assemblies, recent experiments demonstrate that odd-mobile impurities can alter defect proliferation, melting behaviour, and structural relaxation \cite{bililign2022motile, vyas2026two}. Colloidal chiral crystals, both in experiments and theory, show that odd mobility introduces phenomena like edge flows at interfaces \cite{soni2019odd, massana2021arrested, nelson2025topological, caporusso2024phase, caprini2025bubble}, effects that are also seen in lattice models of odd systems \cite{wojcik2026chiral} and dense systems of human cells \cite{yashunsky2022chiral}. More generally, odd interactions are known to accelerate convergence in computational schemes of interacting systems \cite{ghimenti2023sampling, ghimenti2024irreversible}. Our results suggest that such relaxation acceleration and enhanced dynamics may emerge naturally from microscopic odd couplings in dense many-body systems, and that already in the absence of self-propulsion, where similar density-inversions have already been reported for (microscopic) active chiral Janus particles \cite{das2024flocking} and 
in (macroscopic) chiral Hexbug\texttrademark\ systems \cite{yaelprivatecommun}.

Odd mobility coefficients are increasingly recognised as a generic feature of active and biological systems in which microscopic chirality and time-reversal symmetry breaking generate antisymmetric transport responses. Phenomena such as bacterial rheotaxis \cite{marcos2012bacterial}, chiral flows in dense suspensions of microswimmers \cite{li2024robust}, and complex navigation strategies of sperm cells \cite{zhang2016human} and microorganisms \cite{perez2019bacteria} are typically attributed to self-propulsion combined with hydrodynamic interactions or boundary effects. Our results highlight that such odd transport coefficients alone, even in the absence of explicit self-propulsion in the coarse-grained description, can substantially modify tracer diffusion and drift. Whether and how such effective odd interactions contribute to transport anomalies in crowded biological media, thus, remains an interesting open question.

While defect-mediated melting, relaxation acceleration, and (biological) edge currents are often discussed separately, a unified theoretical framework connecting these phenomena in dense many-body systems remains largely lacking. Our present work suggests that starting on a system of odd interacting agents could provide a common microscopic ingredient and serve as a basis to model dense systems.

\ack

E. K., A. S., and R. M. 
acknowledge support by the Deutsche Forschungsgemeinschaft (grants No. SPP 2332 - 
492009952, SH 1275/5-1 and ME 1535/22-1).

\appendix

\section{Hard-disk problem}
\label{app_A}

In order to solve the hard-disk problem, we have to find a solution to the following Laplace equation with reflecting boundary condition,

\begin{subequations}\label{eq:A1}
\begin{align}
 &\Delta_x g(\mathbf{x})=0,\ \text{for}\ x\geq 1, \label{A_1_g_x_HSa}\\ 
&\hat{\mathbf{x}} \cdot \boldsymbol{D} \ \mathbf{f}_{\mathrm{ext}}=  - \hat{\mathbf{x}} \cdot  \boldsymbol{D} \nabla_x g(\mathbf{x}),\ \text{at}\ x=1. \label{A_1_g_x_HSb}
\end{align}
\end{subequations}
The general solution in radial coordinates $\mathbf{x}=(x, \theta)$ is of the type 

\begin{equation}
\label{A_1_general_solution}
    g(\mathbf{x})=\sum_{n=-\infty}^{+\infty} x^n (a_n \cos{(n\theta)} + b_n \sin{(n\theta)}).
\end{equation}
The left-hand sided of Eq.~\eqref{A_1_g_x_HSb} is a sum of cosine and sine for $n=1$, which, together with the natural boundary condition at infinity, leads to the ansatz

\begin{equation}
    g(\mathbf{x})= (a\cos{(\theta)} + b \sin{(\theta)})/x = \hat{\mathbf{x}} \cdot \hat{\mathbf{a}}/x.
\end{equation}

with $\hat{\mathbf{x}}=(\cos{\theta}, \sin{\theta})^\mathrm{T}$. Using this ansatz, the gradient of $g(\mathbf{x})$ can be expressed in polar coordinates as $\nabla_x g(\mathbf{x})=[\hat{\boldsymbol{\theta}}(\hat{\boldsymbol{\theta}}\cdot \hat{\mathbf{a}})-\hat{\mathbf{x}}(\hat{\mathbf{x}}\cdot\hat{\mathbf{a}})]/x^2$ where $\hat{\boldsymbol{\theta}}$ is the angular unit vector,  $\hat{\boldsymbol{\theta}}=\{-\sin{\theta}, \cos{\theta}\}$. Using the Levi Civita symbol $\boldsymbol{\epsilon}$, it is possible to relate the latter to the polar unit vectors as $\hat{\boldsymbol{\theta}}=-\boldsymbol{\epsilon}\, \hat{\mathbf{x}}$ such that Eq.~\eqref{A_1_g_x_HSb} becomes 

\begin{equation}
\label{A_1_boundary}
   \hat{\mathbf{x}} \cdot \boldsymbol{D} \, \mathbf{f}_{\mathrm{ext}} = \hat{\mathbf{x}} \cdot \boldsymbol{D}^\mathrm{T} \hat{\mathbf{a}}.
\end{equation}
Eq.~\eqref{A_1_boundary} should hold for every value of the angular coordinate $\theta$, which implies that we can deduce the vector equality

\begin{equation}
\label{A_1_vector}
   \boldsymbol{D} \, \mathbf{f}_{\mathrm{ext}} = \boldsymbol{D}^\mathrm{T} \hat{\mathbf{a}}, 
\end{equation}
which is solved for the vector $\hat{\mathbf{a}}$ to give the result reported in Eq.~\eqref{unit_vector_HS} of the main text.

The average interacting force on the tracer particle is defined as
\begin{equation}
\label{A_1_ave_force_int_1}
    \mathbf{f}_\mathrm{int}=-\left<\nabla_1 U_N\{ \mathbf{r}_j\}\right>=-N\left<\nabla_1 U(r)\right>,
\end{equation}
where the average is performed over the relative steady state distribution, and the second equality follows from considering that each tracer-host pair gives an independent contribution to the average interaction force in the dilute limit and when only two-body interactions are considered. Particle two is taken as a representative of the interaction and $r=|\mathbf{r}| = |\mathbf{r}_1 - \mathbf{r}_2|$. The average interaction between the tracer and the host particles when the tracer particle is subjected to an external drift force, thus, is computed from the two-body problem
\begin{equation}
    \label{A_1_ave_force_int_2}
    \left<\nabla_1 U(r)\right> = - \int \mathrm{d}\mathbf{r}\, \,  P_\mathrm{rel}^{\mathrm{ss}}(\mathbf{r}) \nabla_r U(r) = - \sigma \int \mathrm{d}\mathbf{x}\,  P_\mathrm{rel}^{\mathrm{ss}}(\mathbf{x}) \nabla_x U(x),
\end{equation}
where
\begin{equation}
\label{app_ansatz}
P_\mathrm{rel}^{\mathrm{ss}}(\mathbf{x})=\frac{\Theta(x-1)}{V}\left(1+ \mathrm{Pe} \,  \frac{\hat{\mathbf{a}}\cdot\hat{\mathbf{x}}}{x}\right),
\end{equation}
and $\mathbf{x} = \mathbf{r}/\sigma$ is the dimensionless relative coordinate, $\sigma$ denotes the particle diametre, $\hat{\mathbf{x}}= \mathbf{x}/x$ is the unit vector along the particle distance, $\beta=1/k_\mathrm{B}T$ the inverse thermal energy, and $\mathrm{Pe} = \sigma \beta|\mathbf{f}_\mathrm{ext}|/2$ is the P{\'e}clet number, encoding the strength of the external drift.

The equilibrium part of $P_\mathrm{rel}^{\mathrm{ss}}$, i.e., the first term in the ansatz of Eq.~\eqref{app_ansatz}, integrates to zero and the remaining term is evaluated with the help of the relation 
\begin{equation}
    \nabla_x \mathrm{e}^{-\beta U(x)} =- \beta\hat{\mathbf{x}}\, \mathrm{e}^{-\beta U(x)}\, \frac{\mathrm{d}}{\mathrm{d}x}U(x).
\end{equation}
In the case of hard-disk interactions, the Boltzmann distribution corresponds to the Heaviside function and its derivative is considered as a Dirac delta centreed at $x=1$, and we can rewrite the former as \cite{kalz2024oscillatory}
\begin{equation}
    \Theta(x-1)\nabla_xU(x)=-k_\mathrm{B}T\hat{\mathbf{x}}\, \delta(x-1). 
\end{equation}

Inserting this relation into Eq.~\eqref{A_1_ave_force_int_2}, we obtain for the average interaction force
\begin{equation}
\label{A_1_ave_force_int_4}
    \mathbf{f}_\mathrm{int} =  -\frac{N}{V} \frac{\sigma^2}{2} |\mathbf{f}_\mathrm{ext}|   \int d\mathbf{x} \delta(x-1) \hat{\mathbf{x}} \otimes\hat{\mathbf{x}}\cdot \hat{\mathbf{a}}/x = -\frac{2\phi}{\pi} \int_0^{2\pi} d\theta \, \hat{\mathbf{x}} \otimes \hat{\mathbf{x}} \left[ a_{\parallel} \mathbf{1} + a_{\perp} \boldsymbol{\epsilon} \right] \cdot \mathbf{f}_\mathrm{ext}.
\end{equation}
The geometric part of the integral can be evaluated using 
$\int_0^{2\pi} \mathrm{d}\theta \, \hat{\mathbf{x}}\otimes \hat{\mathbf{x}}= \pi\boldsymbol{1}$, which yields the result reported in Eq.~\eqref{ave_inter_force_1} of the main part.

\section{Short-range attraction problem}
\label{app_B}

The solution in the presence of the piecewise constant attractive potential sketched in Fig.~\ref{fig:_attractive_potential} of the main part requires that the nonequilibrium radial distribution function is defined piecewise. In each region $g(\mathbf{x})$ should be a solution of the Laplace equation and connected via an appropriate boundary condition through the discontinuous jumps in the potential. In particular, in the attractive region, $1 \leq x < l+1$, $g(\mathbf{x})$ should satisfy in $x=1$ the reflecting boundary condition Eq.~\eqref{A_1_g_x_HSb}, but different from the hard-disk case, the region is of finite size $l$. The consequence of the latter condition is that we should also consider unbounded solutions. For these reasons Eq.~\eqref{A_1_general_solution} should be restricted to $n=1,-1$ which can be expressed in radial coordinates as

\begin{equation}
\label{B_1_eq_atttractive_well}
    g(\mathbf{x})= \mathbf{b} \cdot \mathbf{x} + (\mathbf{c} \cdot \hat{\mathbf{x}})/x,
\end{equation}
where $\mathbf{b}$ and $\mathbf{c}$ are unknown constants. They are related by the reflecting boundary condition at $x=1$, which, once evaluated, results in
\begin{equation}\label{B_1_reflect}
\mathbf{c}=\frac{\boldsymbol{D}^2}{1+\kappa^2}\left[ \mathbf{b}+\mathbf{e}_{f} \right].
\end{equation}

The outer region, $x> l+1$, instead is unbounded, and for this reason, the solution should decay to zero at infinite distance. In addition, we must ensure the continuity of the function $g(\mathbf{x})$ at the value $x=l+1$ so that there is a well-defined value of the average interaction force. The last two requirements lead to 
\begin{equation}
\label{B_2_outer}
    g(\mathbf{x})= (\mathbf{d} \cdot \hat{\mathbf{x}})/x,
\end{equation}
where $\mathbf{d}$ is an unknown. The continuity of the function at $x=l+1$ translates into the following relation between the unknowns $\mathbf{b}$, $\mathbf{c}$ and $\mathbf{d}$

\begin{equation}\label{B_1_continuity}
\begin{split}
(l+1)^2 \mathbf{b} + \mathbf{c} = \mathbf{d}.
\end{split}
\end{equation}

The final equation to close the problem arises from imposing that the relative particle flux $\mathbf{J}_r$ is continuous in the radial direction at the value $x=l+1$. This condition ensures that the number of particles is conserved across the discontinuity of the potential. A similar argument as that leading to Eq.~\eqref{A_1_vector} in the solution for the hard-disk case, eventually leads to

\begin{equation}\label{B_1_flux_radial}
\begin{split}
 e^y   \left[\boldsymbol{D} (\hat{\mathbf{e}}_{f}-\mathbf{b}) -\frac{1+\kappa^2}{(l+1)^2}\boldsymbol{D}^{-1}\mathbf{c} \right] =  \left[\boldsymbol{D} \ \hat{\mathbf{e}}_{f}- \frac{1+\kappa^2}{(l+1)^2}\boldsymbol{D}^{-1}\mathbf{d}  \right] 
\end{split}
\end{equation}

Equations~\eqref{B_1_continuity}, \eqref{B_1_flux_radial} and \eqref{B_1_reflect} form a closed set of equations and can be jointly solved to find the unknown $\mathbf{b}$ as
\begin{equation}
\label{B_1_a}
\begin{split}
\mathbf{b}&=(1-e^y)((l+1)^2-1) \left[(1+\kappa^2)\boldsymbol{D}^{-2} + (e^y(l+1)^2 +1-e^y)\mathbf{1} \right]^{-1} \hat{\mathbf{e}}_{f}\\
&= D_0\frac{(1-e^y)(L^2-1)}{((L^2+1)+e^y(L^2-1))^2+\kappa^2(1-L^2)^2(1-e^y)^2} \\
& \quad \times \left[\mathbf{1}+\frac{2\kappa L^2}{(L^2+1)+e^y(L^2-1)+\kappa^2(1-L^2)(1-e^y)}\boldsymbol{\epsilon}\right]\hat{\mathbf{e}}_{f}
\end{split}
\end{equation}

where $L=(l+1)$. The latter can be expanded in first order in $y$ (small attraction), which yields
\begin{equation}
\label{B_1_a_exp}
    \mathbf{b} = - y \frac{(l+1)^2-1}{2(l+1)^2}\boldsymbol{D}\, \hat{\mathbf{e}}_f +O(y^2).
\end{equation}

Employing similar arguments as in Appendix~\ref{app_A}, the average interacting force now can be evaluated from the knowledge of the radial distribution function: the equilibrium part is a summation of Heaviside functions and the integral in the radial coordinates reduce to the evaluation of two Dirac delta functions at $x=1$ and $x=l+1$ 
\begin{equation}
\label{B_1_f_int_1}
       \mathbf{f}_\mathrm{int}=-\frac{N}{V} \frac{\sigma^2}{2} |\mathbf{f}_\mathrm{ext}|   \int d\mathbf{x} \, g(\mathbf{x}) \hat{\mathbf{x}} \, \left[e^{y} \delta(x-1)+(1-e^{y})\delta(x-(l+1))\right].
\end{equation}

As we imposed the continuity of $g(\mathbf{x})$ at $x=l+1$, the average interaction force is well-defined and the integral can be evaluated to yield

\begin{equation}
\label{B_1_f_int_2}
    \mathbf{f}_\mathrm{int}=-2\phi \left[ \mathbf{c} +  \mathbf{b}((1-e^y)(l+1)^2 + e^y)\right]|\mathbf{f}_\mathrm{ext}|.
\end{equation}
If we expand to linear order in $y$, we obtain the result of Eq.~\eqref{effective_mobility_micro_att} in the main part.

\bibliography{shorttitles,bibliography_Abhi}

\end{document}